# In-Situ Self-Monitoring of Real-Time Photovoltaic Degradation Only Using Maximum Power Point – the Suns-Vmp Method

Xingshu Sun,[1] Raghu Vamsi Krishna Chavali,[1] and Muhammad Ashraful Alam[1,*]

[1] Network of Photovoltaic Technology, Purdue University, West Lafayette, IN, 47907, USA

*Corresponding author: alam@purdue.edu

*Abstract* — **The uncertainties associated with technology- and geography-specific degradation rates make it difficult to calculate the levelized cost of energy (LCOE), and thus the economic viability of solar energy. In this regard, millions of fielded photovoltaic (PV) modules may serve as a global testbed, where we can interpret the routinely collected maximum power point (MPP) time-series data to assess the time-dependent "health" thereof. The existing characterization methods, however, cannot effectively mine/decode these datasets to identify various degradation pathways of the corresponding solar modules. In this paper, we propose a new methodology, i.e., the Suns-Vmp method, which offers a simple and powerful approach to monitoring and diagnosing time-dependent degradation of solar modules by physically mining the MPP data. The algorithm reconstructs "IV" curves by using the natural illumination- and temperature-dependent daily MPP characteristics as constraints to fit the physics-based compact model. These synthetic IV characteristics are then used to determine the time-dependent evolution of circuit parameters (e.g., series resistance) which in-turn allows one to deduce the dominant degradation mode (e.g., corrosion) for the modules. The proposed method has been applied to analyze the MPP data from a test facility at the National Renewable Energy Laboratory (NREL). Our analysis indicates that the solar modules degraded at a rate of ~0.7 %/year due to discoloration and weakened solder bonds. These conclusions are independently validated by outdoor IV measurement and on-site imaging characterization. Integrated with physics-based degradation models or machine learning algorithms, the method can also serve to predict the lifetime of PV systems.**

## I. INTRODUCTION

As an alternative renewable energy resource, photovoltaics (PV) has experienced exponential growth over the last several decades. For investors, an important metric to benchmark the financial viability of PV against other energy resources is the levelized cost of electricity (LCOE). However, the current estimates of the LCOE for PV often rely on the presumption of a linear performance degradation over time. Unfortunately, this presumption leads to an inaccurate LCOE because PV degradations are inherently nonlinear [1]. Also, the rate and magnitude of PV degradation depends sensitively on cell technology and vary substantially across geographic locations

[2], [3]. Hence, a "technology-agonistic" in-situ monitoring method – that characterizes the temporal PV degradation in real time while taking the meteorological information into account – can improve our understanding of the technology- and location-specific degradation rates. This will improve LCOE estimates and suggest opportunities for reliability-aware technology improvement.

There have been many studies on PV reliability reported in the literature, based on different characterization methodologies. These methodologies can be roughly divided into two groups: off-line and on-line techniques.

Typical off-line techniques examine the temporal degradation by periodically and temporarily disconnecting the solar modules for a detailed characterization. For instance, Jordan *et al.* [4] and Sutterluetti *et al.* [5] inspected the degradation mechanisms of PV systems by interpreting IV curves based on the physically-defined five parameter model and the empirical loss factors model (LFM), respectively. Additional sophisticated characterization techniques (e.g., electroluminescence and infrared imaging) can even yield the spatial-resolved degradation analysis for fielded solar modules [6], [7]. Indeed, these off-line methods are incredibly powerful for degradation characterization; however, they require interrupting the normal operation of solar modules at the maximum power point, hence not suitable for continuous monitoring.

On-line techniques, on the other hand, rely on information routinely collected from solar modules. For example, Refs. [8], [9] have analyzed the on-line temporal evolution of PV degradation by continuously examining three time-series performance metrics: (a) $DC/G_{POA}$, the ratio of DC power over the plane-of-array irradiance [10], (b) the performance ratio (PR), a number between 0 and 1 (under STC conditions) equal to the ratio between actual energy yield and nameplate rating [11], (c) the regression PVUSE method that empirically translates on-site output power to the standard test condition (STC) [12]. These methods have the advantage that the modules are not disconnected/interrupted for characterization. The understanding of the degradation pathways, which is critical to establishing the fundamental physics of degradation and promoting reliability-aware design, is still missing from these analysis.

Another on-line characterization approach involves PV data analyzed by statistical machine learning algorithms [13]–[15]. Machine Learning has been proved to be a potent tool to analyze massive data and generate useful insights for different applications. Nonetheless, the weight functions in these algorithms are not physically defined, and it can be difficult to correlate the weights to specific degradation mechanisms. Moreover, network training necessitates a tremendous amount of field data spanning across different geographic locations and technologies as training sets, which are not easily accessible. Therefore, an on-line methodology that can physically and continuously track the degradation of PV systems in real time by interpreting the available field data (and providing insights obtainable only by off-line techniques) can be a transformative tool for the PV community.

Inspired by the well-known Suns-Voc method, in this paper we have developed a simple and powerful strategy to mine the time-series field data to yield a deep understanding of PV reliability and identify various degradation pathways. The Suns-Voc method [16], where one monitors the open-circuit voltage by manually varying illumination intensity of a solar simulator (see. Fig. 1), has been demonstrated to be a useful characterization tool during module development. Obviously, it cannot apply directly to field data composed exclusively of maximum power point (MPP) current ($I_{mp}$) and voltage ($V_{mp}$) information. Hence, we propose **the Suns-Vmp method** that, by taking advantage of the natural daily variation of sunlight, can deduce circuit parameters as a function time by fitting the reconstructed MPP "IV" throughout the day, see Fig. 1. By systematically and physically mining the streaming MPP data, the method can monitor the reliability of solar modules in real time.

In this paper, we begin by introducing the detailed methodology of the Suns-Vmp method in Sec. II. In Sec. III, the Suns-Vmp method is applied to an NREL test facility to extract the degradation rate and the dominant degradation modes. Sec. IV discusses the implication of the Suns-Vmp method on the prediction and design of PV reliability and the limitation herein. Finally, we summarize the paper in Sec. V.

## II. THE SUNS-VMP METHOD

In this section, we will discuss the Suns-Vmp algorithm, as summarized in Fig. 2. The algorithm has the following four steps: 1) develop the physics-based equivalent circuit model for a specific technology; 2) extract pristine (time-zero) circuit parameters based on datasheet/pre-installation IV characteristics; 3) preprocess MPP data to reconstruct IV characteristics synthetically, and 4) finally, analyze the time-degradation of circuit parameters for insights regarding the dominant degradation modes.

### A. Step 1: Development and Choice of the Equivalent Circuit (Compact Model)

Mainstream PV technologies can be categorized into three groups: 1) p-n homojunction (e.g., c-Si and GaAs), 2) p-i-n junction (e.g., a-Si and perovskites), and 3) p-n heterojunction (e.g., CIGS and CdTe). Depending on a particular technology, we select the corresponding equivalent circuit in the Suns-Vmp method, see for example, [17] for CIGS, [18] for perovskites, [19] for silicon heterojunction. *Since a solar cell is exposed to varying illumination intensity and temperature, the equivalent circuit must be capable of describing the illumination- and temperature-dependent IV curves.*

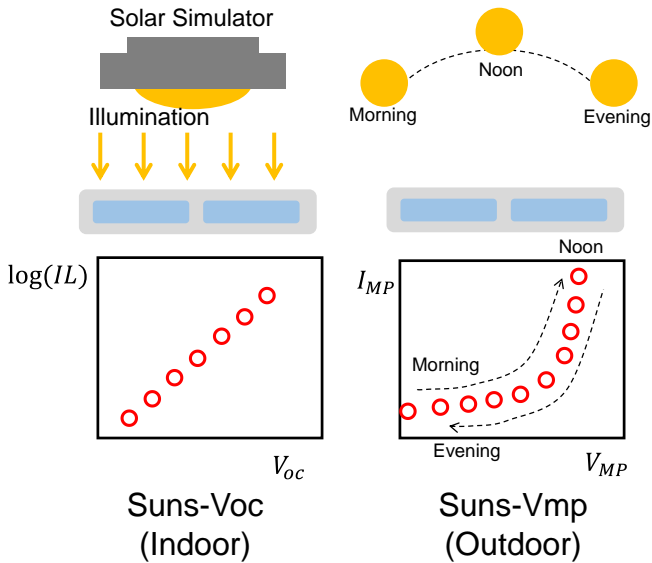

Fig. 1 A schematic illustration to explain the working principles of the Suns-Voc and Suns-Vmp method.

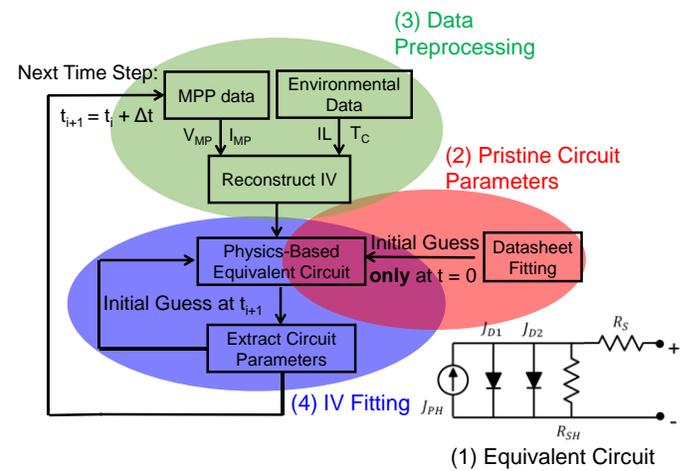

Fig. 2 The flowchart of the Suns-Vmp method. The analytical formulation of the five-parameter model is from [20], [33] and summarized in the supplementary material.

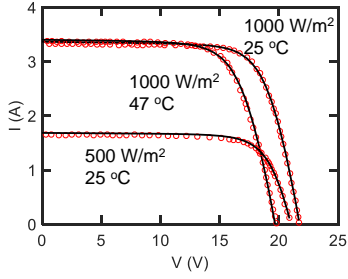

Fig. 3 Initial fitting to the datasheet (Siemens M55 [30]) for time-zero circuit parameters. The extracted circuit parameters are summarized in the supplementary material.

In this paper, we will demonstrate the Suns-Vmp method on a c-Si PV system, therefore we make use of the well-known five parameter model for Si solar modules [20], which explicitly accounts for the illumination- and temperature-dependencies of circuit parameters, namely, $J_{PH}$, $J_{01}$, $J_{02}$, $R_{SH}$, and $R_S$, see Fig. 2. The complete set of equations and parameter descriptions for the five parameters is summarized in the supplementary material (SI). If needed, the five parameter model can be generalized to include nonlinear shunt resistance [21] and temperature- and illumination-dependent series resistance [22], [23].

### B. Step 2: Extracting Pristine Module Parameters

Next, we extract the pristine (time-zero) module parameters (before the module is fielded) as robust initial guesses for the Suns-Vmp method. We do so by fitting the complete illumination- and temperature-dependent IV measurements available from the datasheet or pre-installation measurement. With the robust initial guesses, we can eliminate multiple solutions in the sequential IV fitting process, see Fig. 3. Typical datasheet usually provides a set of full IV measurement under various illumination and temperature conditions which guarantee the uniqueness of the extracted circuit parameters and consequently the robustness of the initial guess.

### C. Step 3: Preprocessing MPP Data

After obtaining the time-zero circuit parameters, we construct – at any time during the onsite operation – a synthetic IV curve by sampling MPP data over a given period (typically 2-3 days, referred as measurement window hereafter). Recall that in the Suns-Voc measurement [24], [25], one traces the open circuit voltage of solar cells, through deliberately varying the intensity of the solar simulator, to construct the IV curve in the absence of series resistance. In the Suns-Vmp method, however, we take advantage of the natural temporal variation of the sunlight (the plane-of-array irradiance: $G_{POA}$) and the cell temperature ($T_C$) to track the maximum power point. Hence, due to the changing $G_{POA}$ and $T_C$, the module output $I_{mp}$ and $V_{mp}$ (operating current and voltage at the maximum power point, respectively) increase from morning to noon then decrease from noon to evening, see Fig. 4(a). For example, if the data is recorded every 10 minutes of 8 diurnal hours over a

3-day measurement window, then 144 data points of four variables (i.e. $G_{POA}$, $T_C$, $I_{mp}$, $V_{mp}$) are available to calculate the circuit parameters of the compact model, namely, calibrating the circuit parameters until the MPP IV is reproduced as shown in Fig. 4(b). *Note that Suns-Vmp method does not interrupt the normal module operation by disconnecting solar modules for IV sweeps or deviating them from the MPP bias [16], [26]; thus the technique empowers characterization of solar modules in real-time operation.*

In the Suns-Vmp methodology, to reduce uncertainties in the extraction, we also explicitly preprocess the data to account for 1) cell-to-module temperature difference, 2) spectral mismatch between pyranometer and solar modules, and 3) reflection loss as a function of time. The specific steps are summarized in the SI. Also, while the basic algorithm is easy to understand, it is important to realize that the $(G_{POA}, T_C, I_{mp}, V_{mp})$ may involve noisy or corrupted data. In this case, the window duration must be chosen judiciously and the corrupted data must be rejected, for a robust parameter extraction of the compact

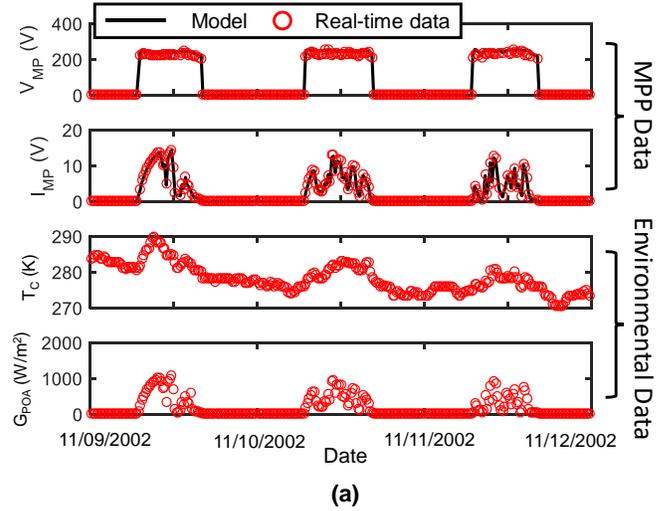

(a)

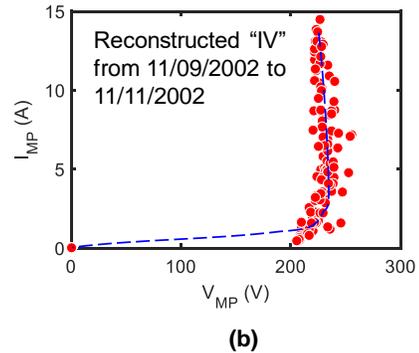

(b)

Fig. 4 (a) Three-day MPP and environmental data (circles) from 11/09/2002 to 11/11/2002 of the test facility in Sec. III. The fitting results of the MPP data (solid lines) using the Suns-Vmp method is also present. (b) An illustration of reconstructing "IV" from the MPP data in (a).

model. Hence, we have developed a physics-based self-filtering algorithm to preprocess the data as follows before fitting (see the SI material for additional details).

The *measurement window* of MPP data must be chosen such that it is long enough to contain sufficient illumination/temperature variations, but short enough such that the module does not degrade significantly within the window. The time-scale of degradation processes is slow [13], thereby the circuit parameters can be assumed to be constant over the course of a few days. Hence, the recommended measurement window of MPP data can be up several days (e.g., three days in Fig. 4), as long as there exists sufficient variation in illumination and temperature to reconstruct the MPP IV. In the case of catastrophic degradation (such as partial shading degradation in thin-film solar modules [7]), the extracted circuit parameters become the average value of pre- and post-degradation values over time.

### D. Step 4: MPP IV Fitting Algorithm:

After reconstructing MPP IV and preprocessing environmental data, we proceed with using rigorous fitting algorithms to model the measured MPP data and extract circuit parameters. In this paper, we have used the nonlinear least-squares fitting algorithm and bio-inspired particle swarm optimization (PSO) (i.e., "lsqcurvefit" and "particleswarm" functions in Matlab® [27], respectively), both of which have been found to give identical results. Note that both fitting algorithms require a lower and upper bound of each circuit parameter at each time step. In our analysis, circuit parameters are assumed to degrade monotonically as a function of time (i.e., no recovery) with a *maximum* degradation rate of 1 %/day, except for the short-circuit current $J_{Ph}$. Hence, given the used length of measurement window, the upper and lower bound can be determined. Since the short-circuit current may fluctuate abruptly due to soiling and precipitation, the upper and lower bound thereof are set to be the datasheet short-circuit current and zero, respectively. Even though recovery of certain degradation pathways is possible (e.g., output power recovers after removing voltage stress for potential induced degradation [28], [29]), such recovery is expected to be negligible due to constant environmental stress (e.g., thermal cycling, moisture exposure) applied on the operating solar modules.

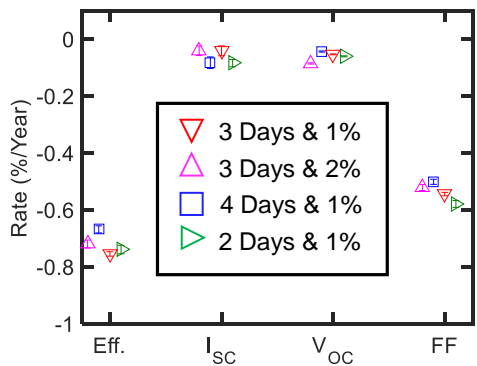

Fig. 5  Degradation rate of performance metrics of the negative array as a function of different settings (i.e., measurement window and maximum degradation rate of circuit parameters) in the Suns-Vmp method.

For any inverse algorithm such as the Suns-Vmp method, one must ensure that the uniqueness of the degradation analysis. Hence, we present a sensitivity analysis of these two algorithm parameters, i.e., measurement window and maximum degradation rate of circuit parameters, on the final extraction of degradation rates, see Fig. 5. Our results show that moderate change in the algorithm parameters in the Suns-Vmp method does not interfere with the final results – the deduced degradation rates of performance metric remain unique.

In the next section, we will demonstrate the Suns-Vmp method on an NREL test facility with recorded field data to analyze the degradation of solar modules in real time. The analysis will reveal the possible root causes of power losses by physically interoperating the time-dependent circuit parameters.

## III. APPLICATION TO FIELD DATA

### A. Introduction to Field Data

The studied PV system (No: NREL x-Si #7) perches at the west side of the Solar Energy Research Facility (SERF) building at the National Renewable Energy Laboratory (NREL), Golden, CO, USA. It comprises two arrays with negative and positive monopole, each of which consists of five strings with 14 x-Si Siemens M55 solar modules [30] totaling to around 7.42 kW capacity. In 2007, a negatively grounded inverter replaced the previous bipolar inverter, but we maintain the bipolar naming convention (negative versus positive) in this paper. The modules are 45º tilted and oriented 22º east of south. All the onsite MPP and environmental data (illumination and module temperature) including the metadata were retrieved from the publicly accessible NREL PV Data Acquisition (PVDAQ) database [31] with time resolution spanning from 1 min to 15 min. The analyzed field data is from 05/13/1994 to 12/31/2014. Three measurements of module temperature were initially recorded by thermocouples attached to the backsheets but significantly inconsistency was found after the eighth year. Therefore, we applied the calibrated Faiman model [32] to obtain module temperature. In addition to continuous MPP data, outdoor IV measurements were also carried out at the array level using a portable Daystar I–V tracer. These IV data sets help us validate the analysis obtained from the Suns-Vmp method. More details on this PV systems can be found in SI.

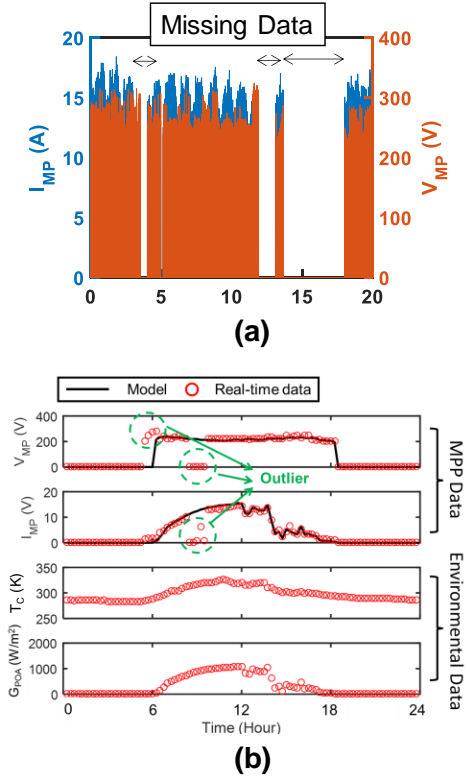

**Fig. 6** (a) 20-year data of $I_{MP}$ and $V_{MP}$ of the negative monopole. (b) One-day data exhibits the existence of corrupted outlier points.

Figure 6 displays the example data while highlighting the two major challenges of analyzing this field data – 1) several gaps even up to 5 years of absent field data and 2) corrupted data with outliers possibly due to instrumentation error, inverter clipping, weather condition, etc. First, to mitigate the uncertainty in deducing the circuit parameters induced by missing data, the Suns-Vmp method makes use of the results from the previous time step as initial guesses and establishes the upper/lower bounds with a preset maximum change rate when fitting the MPP IV. Second, we need a self-consistent scheme to detect and remove these outliers. Toward this goal, we have created a continuous self-filtering algorithm as summarized in the SI. Enabled by these techniques, the Suns-Vmp method can retain excellent error control, *i.e., the mean absolute percentage error (MAPE) is less than 5% for both Vmp and Imp throughout the entire 20-year analysis.*

### B. Results and Validation

Figure 7 summarizes the extracted circuit parameters of the negative array by fitting the five-parameter model (see Fig. 2) in [20], [33] to the MPP data with a three-day measurement window over a span of 20 years (from 1994 to 2014). The positive array also shows a very similar result, therefore not included here. The maximum photocurrent ($J_{PH}$) fluctuates possibly due to the accumulation of dust/snow [34] or recalibration of the pyranometer during 20 years. However, it is expected that this fluctuation in $J_{PH}$ does not disturb the extraction of other parameters, since the five-parameter model assumes voltage-independent $J_{PH}$ and therefore the fluctuation will just shift the IV in Fig. 4 but not change the underlying IV characteristics (shape). Remarkably, it appears that all the circuit parameters in Fig. 7 were degrading (e.g., shunt resistance ($R_{SH}$) reduces, and series resistance ($R_S$) increases). To quantify the degradation rate, we calculate the efficiency at standard test condition (STC) at each time step, see Fig. 8.

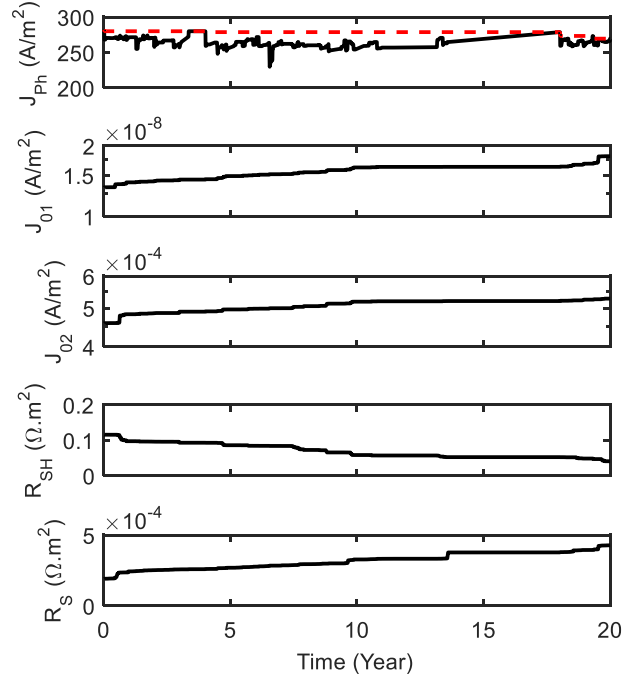

**Fig. 7** The extracted circuit parameters under standard test condition of the five-parameter model for the negative array as a function of time. Notations: $J_{PH}$ is the maximum photocurrent density; $J_{01}$ is the reverse saturation current density with ideality factor of 1; $J_{02}$ is the reverse saturation current density with ideality factor of 2; $R_{SH}$ is the shunt resistance; $R_S$ is the series resistance. $J_{PH}$ is corrected so that it monotonically decreases with time (red dashed line).

*Validation 1: Comparison to DC/$G_{POA}$ method.* Remarkably, the extracted STC efficiency by the Suns-Vmp method compares well with that of the conventional DC/$G_{POA}$ method [35], showing both the negative and positive arrays near their warranty lifetime (80% of initial efficiency). However, the result obtained from the DC/$G_{POA}$ method shows greater fluctuation than the Suns-Vmp method due to 1) the empirical approaches to filtering outliers and 2) linear temperature-correction of real-time output power to STC by a constant temperature coefficient (which changes over time). Because the Suns-Vmp method uses a physics-based equivalent circuit for outlier filtering and temperature correction, the fluctuation is substantially reduced. Note that, for the Suns-Vmp method, we correct $J_{PH}$ so that it monotonically decreases with time (i.e., soiling loss is recoverable) when calculating the STC efficiency, see Fig. 7.

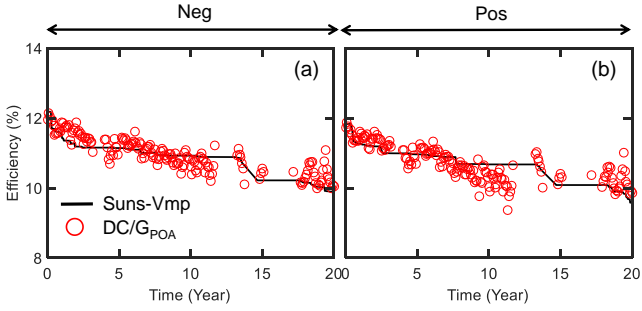

Fig. 8 Temporal STC efficiencies calculated by the Suns-Vmp and DC/G_POA methods for the arrays with a negative (a) and positive monopole (b), respectively.

*Validation 2: Outdoor IV Measurement.* To further validate the Suns-Vmp method, we benchmark the obtained results against those characterized by the periodic outdoor IV measurement through 20 years. Figure 9 shows the comparison between real-time (not STC) PV performance metrics calculated by circuit parameters deduced by the Suns-Vmp and direct outdoor IV measurements. Indeed, we find great consistencies (less than 4% MAPE) between these two methods, which corroborates the accuracy of the extraction by the Suns-Vmp method.

*Validation 3: Parameter degradation Rates.* Besides the performance metric, we also benchmark the rate of change of the performance metrics estimated from the Suns-Vmp method against outdoor IV from [36] in Fig. 10 (top), which again are in good agreement. The error bars are calculated within 95% confidence interval. The degradation rate of the efficiencies for both the negative and positive arrays are around 0.7%/Year. It is noteworthy that the efficiency degradation may be primarily attributed to the reduction in fill factor (-0.6 to -0.4 %/Year), while Voc and Isc only worsen slightly.

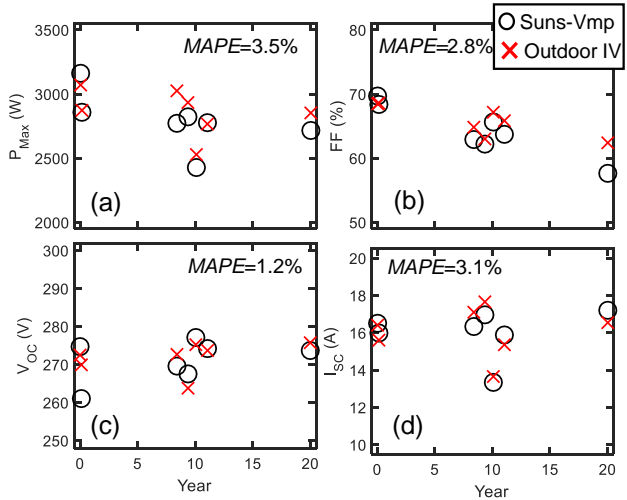

Fig. 9 Comparison of performance metric generated by the Suns-Vmp method and outdoor array IV measurement for the negative array. The mean absolute percentage errors (MAPE) are also labelled in each plot.

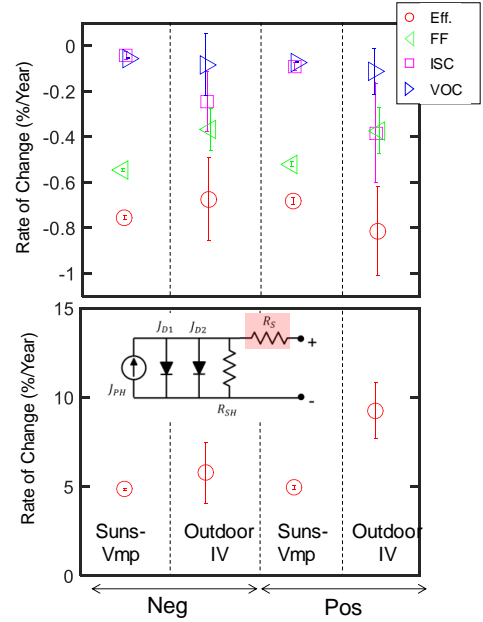

Fig. 10 Rate of change of the performance metrics (top) and series resistance $R_S$ (bottom) of the analyzed PV systems via the Suns-Vmp method and outdoor IV measurement.

We attribute this degradation to the increased series resistance, which erodes fill factor without substantially affecting Voc and Isc. Both the Suns-Vmp and outdoor IV measurement reveal the rapid increment of series resistance at the rate of 5 – 10%/year as shown in Fig. 10 (bottom), which confirms our conjecture of series-resistance induced efficiency degradation.

*Validation 4: Onsite inspection.* Next, we will deconvolve and quantify the power losses ascribed to each circuit parameter to identify the predominant physical degradation pathways. As shown in Fig. 11 (a), we deconvolve the power losses associated with each parameter for the negative array. The key observations are threefold:

1) At the end of 20 years, Fig. 11 (a) elucidates that the increased series resistance is the dominant contributor to efficiency reduction for both the negative and positive polarities. Remarkably, the on-site infrared image in Fig. 11 (b) exhibits localized hot spots caused by solder bond failure, in accord with our deconvolution analysis of increasing series resistance. It is generally known that the failure of solder bonds is because of thermal stress induced by the different thermal expansion coefficients of solder joints and components during repeated thermal cycles [37], [38]. Therefore, solder bonds fail

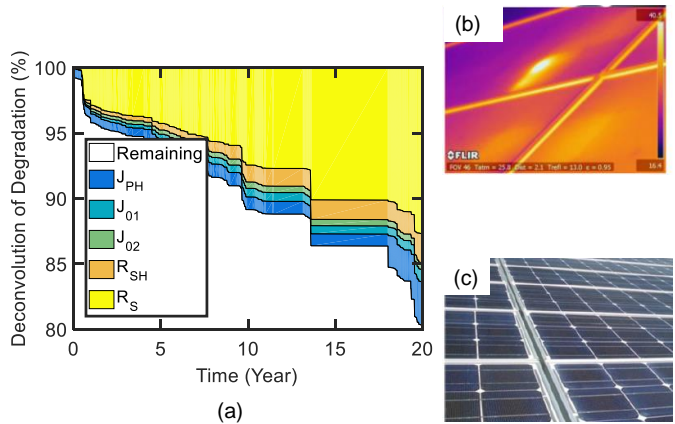

Fig. 11 (a) Temporal degradation deconvolution with respect to circuit parameters for the negative polarity. (b) IR image shows a hot spot caused by solder bond failure. (c) Picture shows that most cells suffer from discoloration in the center. *(b) and (c) are obtained from [36].

(crack) at the cycle of failure in a stepwise fashion [39]. Indeed, the incremental time signature of the series resistance is stepwise in the Suns-Vmp analysis, see Fig. 11 (a).

2) Discoloration of the encapsulants can be expected because of the relatively high ultraviolet light concentration at Denver (altitude of ~1800 m) [40]. Indeed, a photograph of the solar modules in the field shows that the majority of the solar cells suffer from discoloration, see Fig. 11 (c). Meanwhile, notwithstanding the $J_{PH}$ fluctuation shown in Fig. 7, our deconvolution results also manifests a symmetric decrease of $J_{PH}$ and ascribes a significant amount of power loss (~4%) to $J_{PH}$ reduction, an indicator of discoloration. This agreement again confirms the PV degradation diagnosed by the Suns-Vmp method. It is noteworthy that the photocurrent reduction due to discoloration has occurred within the first year of installation. Another study has also found early advent of discoloration, i.e., discoloration has been seen in 50% of the solar module less than five years old [41].

3) The operating voltage of the modules is only around 200 V; therefore, the efficiency degradation by potential-induced degradation (PID) is expected to be insignificant [42]. Indeed, our result confirms this conjecture by showing that only ~3% power loss is due to shunting ($R_{SH}$) and increased recombination currents ($J_{01}$ and $J_{02}$), both of which are effective indicators for PID [43], [44].

As demonstrated here, the Suns-Vmp allows us to quantitatively and qualitatively diagnose the pathology of degraded solar modules exposed in the field by analyzing and interpreting the time signature of individual circuit parameters. All the results have been validated by both outdoor IV measurement and on-site characterization.

## IV. DISCUSSION

In the previous section, we have applied the Suns-Vmp method to an NREL test facility and demonstrated its capability of analyzing the degradation of solar modules in real time. Next, we discuss the potential use of the time-dependent parameters obtained through the analysis and limitations of the approach.

### A. Geography and technology-specific reliability-aware design

The underlying physical degradation mechanisms of PV are strongly contingent on local meteorological factors and different technologies, e.g., solar modules exposed in humid regions are more susceptible to contact corrosion [45], and monolithic thin-film solar modules are vulnerable to partial shading degradation [7]. Similarly, modules more likely to suffer from PID should adopt Corning® Willow™ Glass to impede ion migration [46]. Therefore, ideally, module design ought to be geography- and technology-dependent. However, solar modules are often overdesigned for reliability (perhaps at a considerable cost) so that they can survive a broad range of weather conditions. This is due to the lack of comprehensive understanding of local degradation. The Suns-Vmp method offers an opportunity to efficiently diagnose the degradation pathways of fielded solar modules of different technologies across the entire world. The results can be ultimately collected in a global database, allowing the manufacturers to design and produce the next generation reliable-aware PV with maximized durability.

### B. More accurate long-term reliability prediction

Accurate prediction of long-term energy production by PV systems is crucial to evaluating the bankability thereof. Various degradation pathways depend nonlinearly on stress time and local stress factors (irradiance, voltage, moisture, temperature). Therefore, it is difficult to predict future energy yields based on empirical linear degradation models [2]. In this regard, the Suns-Vmp method can facilitate accurate reliability prediction. Recently, several physics-based degradation models have been developed that can directly map various PV degradation modes (e.g., corrosion, PID, yellowing) to the temporal behavior of circuit parameters [47], [48]. The extracted circuit parameters by the Suns-Vmp method can be used to calibrate these degradation models (e.g., moisture diffusion coefficient for corrosion). Integrated with the weather forecast, the calibrated degradation models will predict the lifespan of solar modules. Alternatively, the time-dependent circuit parameters can train machine learning algorithms; the trained machine learning algorithms [13] can predict PV lifetime. The validity of these predictive approaches, however, remains an interesting open question and requires more rigorous research efforts.

### C. Guidance for collection of field data

The Suns-Vmp methodology highlights the importance of physics-based modeling in creating databases. For example, we have seen fitting of the pristine module characteristics requires temperature- and illumination-dependent IV measurement to ensure a robust and unique initial guess. Second, we have noted that weather data may be corrupted or missing. Thus it is important for PV databases to contain complementary information from multiple sources [49]. Finally, compact model parameters offer an important recipe for improving data compression and computational efficiency; the model parameters can diagnose the module by only deciphering the stored Vmp-Imp information (a byproduct data of normal operation at maximum power point) for the entire duration. This eliminates the need for deliberate measurement of massive IV data [15] and time-consuming collection of field data [41].

### D. Intra-string variability

Process-induced variability can lead to performance variation in the cell, module, or array levels [19], [50], [51], especially for the thin-film PV where binning is not possible. Similarly, various degradation modes introduce local variability as well. For example, non-uniform degradation (e.g., cells adjacent to module edges are more prone to contact corrosion than those located away from the edges [52]; solar modules close to the negative array are more susceptible to PID [53]), etc. As implemented, the Suns-Vmp method uses a single equivalent circuit to analyze a string consisting of multiple modules and thus accounts for "average" variability/degradation. As a result, it is critical to investigate how performance variability can potentially affect the accuracy of the Suns-Vmp method. Therefore, we have tested Suns-Vmp under various scenarios of performance variability, and the results are listed in the SI. Remarkably, our findings highlight that the circuit parameters extracted by the Suns-Vmp method are still valid to interpret PV degradation with *moderate* non-uniformity. Affected by severe non-uniformity, however, the Suns-Vmp method may not be able to identify the primary circuit parameters contributing to power losses. For instance, the Suns-Vmp method could attribute the predominant degradation to the increased recombination current ($J_{01}$ and $J_{02}$), and series resistance $R_S$, whereas the actual degradation is due to reduced shunt resistance $R_{SH}$. For these cases, it will be important to represent the string by a few equivalent circuit models. Despite the increase in the parameter number, the following considerations are expected to simplify the calibration process: 1) availability of time-zero information of each module, 2) the large amount of data available within the measurement window, and 3) several degradation modes (e.g., yellowing) are expected to affect all the modules uniformly, while others (e.g., PID) are dominated by a few modules. Ability to account for non-uniform degradation will be an important direction of future research on this topic.

## V. CONCLUSION

To summarize, we have presented a novel method, i.e., the Suns-Vmp method, for analyzing the PV degradation:

1. The Suns-Vmp method enables in-situ monitoring and diagnosis of PV reliability in real time by systemically and physically mining the time-series MPP data. The method can extract physically defined circuit parameters by fitting IVs consisting of the varying MPP data over a characterization window. The extracted circuit parameters can be used to estimate the STC efficiency, quantitively deconvolute PV degradation, and identify the dominant degradation pathways.

2. We have demonstrated the Suns-Vmp method by analyzing MPP data from an NREL test facility, where physics-based circuit parameters and efficiency of the solar modules have been extracted as a function of time. Independent outdoor IV measurements have systemically validated our results. Our analysis suggests that the PV system degrades at a rate of 0.7%/Year, primarily due to reduced short-circuit current and increased series resistance most likely caused by discoloration and weakened solder bond, respectively. The on-site optical photograph and IR image indeed substantiate our interpretation of the physical degradation pathways, i.e., discoloration and solder bond failure.

3. The analysis of deconvoluting the underlying degradation pathways by the Suns-Vmp method can deepen the current understanding of technology- and geographic-dependent degradation, and inspire more robust environment-specific designs for the next-generation "reliability-aware" solar modules. The Suns-Vmp method can be used to calibrate physics-based degradation models as well as train machine learning algorithms, both of which can then predict power degradation of PV and improve the evaluation of "bankability."

## ACKNOWLEDGMENT

This work was supported by the US-India Partnership to Advance Clean Energy-Research (PACE-R) for the Solar Energy Research Institute for India and the United States (SERIIUS), and the DEEDS program by the National Science Foundation under award #1724728. The authors would like to thank Haejun Chung, Reza Asadpour, and Dr. Mohammad Ryyan Khan for helpful discussion, Dr. Chris Deline and Dr. Dirck Jordan for providing IV measurement, as well as Prof. Mark S. Lundstrom and Prof. Peter Bermel for kind guidance.

# In-Situ Self-Monitoring of Real-Time Photovoltaic Degradation Only Using Maximum Power Point – the Suns-Vmp Method

Xingshu Sun,[1] Raghu Vamsi Krishna Chavali,[1] and Muhammad Ashraful Alam[1]

[1]*Purdue University School of Electrical and Computer Engineering, West Lafayette, IN, 47907, USA.*

## Supplementary Information

## 1. Preprocess Environmental Data

The Suns-Vmp method relies on environment data, i.e., cell temperature and irradiance. The weather information is used as inputs to the equivalent circuit to fit the reconstructed MPP IV. The raw data can contain seasonal irradiance variation and temperature correction. Hence, it is important to preprocess the raw data so that the parameters extracted are accurate and robust. Below, we discuss this issue of data preprocessing in detail.

**Cell Temperature.** Module temperature is typically measured by attaching thermal sensors to the back side of solar modules. The actual cell temperature can be higher than the measured back-side module temperature regardless of convective and radiative heat transfer at the module surfaces. Ref. [1] has developed an empirical equation to predict cell temperature ($T_C$) based on illumination intensity ($G_{POA}$) and module temperature ($T_M$), which is used in this paper.

**Irradiance Data**. In addition to thermal information, we also need the illumination data to perform the Suns-Vmp method. The on-site illumination data is typically measured by pyranometers orientated as same as solar modules to collect the plane-of-array irradiance $G_{POA}$. However, directly applying the raw $G_{POA}$ data to the Suns-Vmp method can cause inaccuracy in extracting short-circuit current because of 1) air mass dependent spectral mismatch between field and standard test condition (STC) and 2) reflection loss of flat-plate solar modules. Thus, one must preprocess $G_{POA}$ data to eliminate the above non-idealities, as discussed below.

**Spectral Mismatch.** The spectral profile of $G_{POA}$ under which MPP data is generated can differ from the AM1.5G spectrum used in the STC for initial rating. Because the extracted circuits from the Suns-Vmp method are eventually corrected to their STC values, the spectral mismatch between real-time field irradiance and STC can contaminate the fitting results primarily for the short-circuit current. Fortunately, the Sandia PV Array Performance Model (SAPM) has developed a polynomial equation to empirically describe the spectral content of solar irradiance as a function of air mass (*AM*) [1]. In this paper, we use the SAPM to correct the real-time $G_{POA}$ to its STC values, where *AM* is calculated by the Sandia PV modeling library [2] and the Direct Normal Incidence (DNI) is retrieved from [3] at the installation location.

**Reflection Loss.** Pyranometers can accept irradiance coming from a highly oblique angle of incidence (AOI) thanks to the doom-shaped glass cover, while flat-plate solar modules are susceptible to reflection loss at high AOI. Consequently, one must also adjust $G_{POA}$ measured by pyranometers to account for reflection loss. In this paper, we also utilize the SAPM module [1] to correct for reflection loss of the direct normal incidence, given the tilt and azimuth angles of the analyzed solar modules.

Although the metrological information is often available from the on-site weather station, this may not be the always the case. In this case, meteorological databases, such as Ref. [3] can be alternative sources for reproducing illumination [2] and temperature information [4].)

## 2. Physics-Based Filtering Algorithm

Outlier data points due to instrumentation error, inverter clipping, weather condition, etc., can exist in the field data [5]. For example, the Imp data point at around 9 am in Fig. S1 shows substantial inconsistency with $G_{POA}$. The inclusion of these outliers in the Suns-Vmp method can induce significant uncertainties in extracting circuit parameters. Therefore, it is necessary to develop a self-consistent scheme to detect and then remove these outliers. Toward this goal, we have created a continuous self-filtering algorithm to eliminate outlier data points, see Fig. S2. The steps are as follows:

1) Fit the MPP data with non-zero POA irradiance using the equivalent circuit (MPP data with zero irradiance always yields zero current and voltage, thereby irrelevant). Note that this fitting step is confined to the MPP data only within the measurement window at a single time step.

2) Calculate the relative error of fitting each MPP data point. If the error is greater than 50%, the corresponding data point is treated as an outlier and discarded.

3) Examine the number of the remaining data points after step 2. If the remaining still consists of more than 80% of the raw data points, proceed to step 4. Otherwise, the corresponding time step is considered as an outlier as a whole (i.e., remove all the data points at this time step), and will not be analyzed further. Rather, the Suns-Vmp method will directly move to the next time-window. The entire measurement window may consist of corrupted data if temporary instrumentations malfunctions for more than a few days.

4) Fit the filtered MPP data by the equivalent circuit and extract the circuit parameters.

5) Move to next time step.

Note that our continuous self-filtering algorithm in this paper has comprehensively accounted for outliers caused by various non-idealities (e.g., cloud brightening, inverter clipping, temperature/illumination stability, pyranometer error); thus, there is no need to create individual data filters as in [5] for the Suns-Vmp method. Moreover, the percentage thresholds in steps 2 and 3 (i.e., 50% for relative error and 80% for the number of remaining data points) is found to work well for analyzing field data, and we do not expect a moderate adjustment of the percentage thresholds will significantly impact the outcome. Enabled by our filtering algorithm, excellent error control has been achieved, i.e., the relative error is less than 5% for both Vmp and Imp through the entire 20-year analysis.

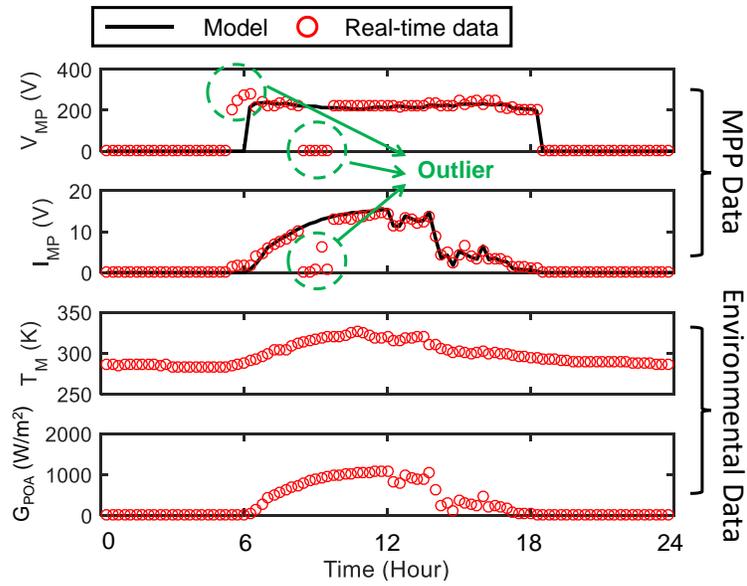

Fig. S1 Raw MPP data with outliers, filtered MPP data, and the environmental data on 05/16/1994 of the NREL test facility.

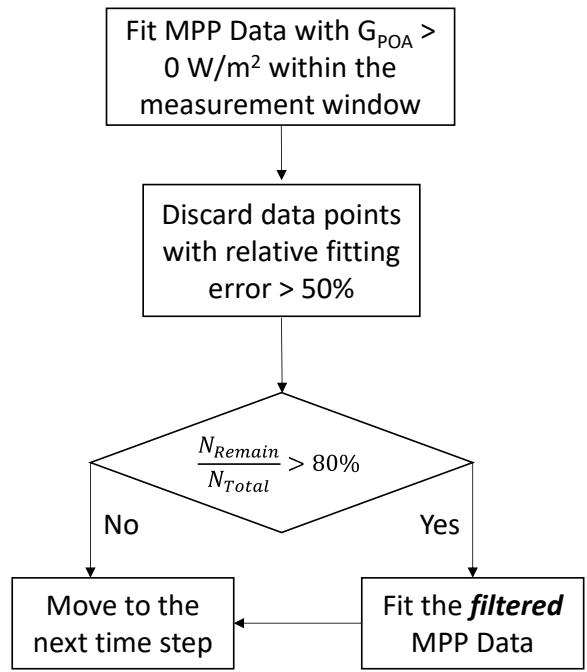

Fig. S2 Flowchart of our self-filtering algorithm to identify and eliminate outlier data points.

## 3. Variability Test of the Suns-Vmp Method

We have tested the Suns-Vmp method under various scenarios of variability using synthetic weather data in Fig. S3. Non-uniform degradation of solar cells in the field can occur due to different degradation pathways and have different levels of non-uniformity. Hence, we have emulated four cases of performance variability: 1) 6 out of 36 cells degrades due to contact corrosion ($R_S$ increases tenfold); 2) 6 out of 36 cells have encapsulant delamination (only retain 80% of initial short-circuit current); 3) 6 out of 36 cells suffer from moderate potential-induced degradation (shunt resistance decrease by one order); 4) 6 out of 36 cells suffer from server potential-induced degradation (shunt resistance decrease by two orders). All the tests of performance variability are summarized in Figs. S4 to S7.

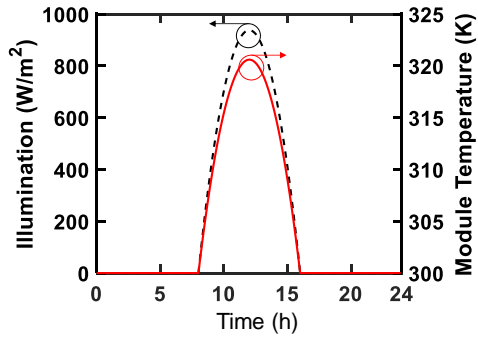

Fig. S3    Synthetic weather data containing hourly illumination and module temperature is used to test the Suns-Vmp method.

As shown in Figs. S3 – S7, the Suns-Vmp method is still capable of diagnosing the pathology of solar modules with non-uniform degradation. For example, the Suns-Vmp method has attributed efficiency degradation to the increased series resistance in Fig. S4. This result, however, is not surprising since series resistance can essentially be aggregated into one single resistance in a series-connected circuit in Fig. S4(a). Remarkably, the Suns-Vmp method is still valid even for non-uniform delamination- and PID-induced degradation where simple superstition of either short-circuit current and shunt resistance of "good" and degraded cells does not hold, see Figs. S5 and S6. The Suns-Vmp, however, cannot correctly extract the degraded circuit parameter by only one single equivalent circuit under severe performance variability, see Fig. S7. Hence, it is recommended to utilize multiple equivalent circuits in the Suns-Vmp method to analyze solar modules with substantial performance variability.

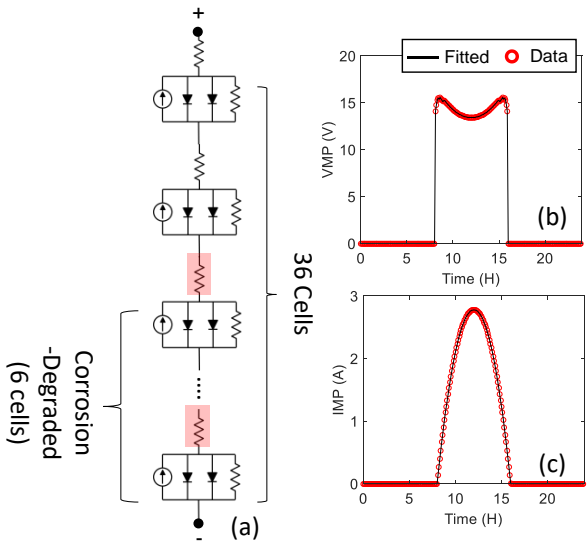

| (d) Extracted parameter by the Suns-Vmp method | | | |
|---|---|---|---|
| | Default (30 cells) | Degraded (6 cells) | Extraction |
| $J_{PH,STC}$ | 282 A/m² | 282 A/m² | 282 A/m² |
| $J_{01,STC}$ | 1.3 x 10⁻⁸ A/m² | 1.3 x 10⁻⁸ A/m² | 1.3 x 10⁻⁸ A/m² |
| $J_{02,STC}$ | 4.6 x 10⁻⁴ A/m² | 4.6 x 10⁻⁴ A/m² | 4.6 x 10⁻⁴ A/m² |
| $R_{Sh,STC}$ | 0.12 Ω.m² | 0.12 Ω.m² | 0.12 Ω.m² |
| $R_S$ | 1.7 x 10⁻⁴ Ω.m² | **1.7 x 10⁻³ Ω.m²** | **4.2 x 10⁻⁴ Ω.m²** |

Fig. S4  (a) A schematic of the simulated 36-cell solar module including 6 cells degraded due to contact corrosion. The degraded circuit elements are also highlighted. (b,c) Vmp and Imp of the solar panel using the synthetic weather data in Fig. A1. Circles are simulated data and solid lines are fitting data using the Suns-Vmp method. (d) Table summarizes input parameters (both default and degraded) and extracted parameter set using the Suns-Vmp method (affected parameters are in bold).

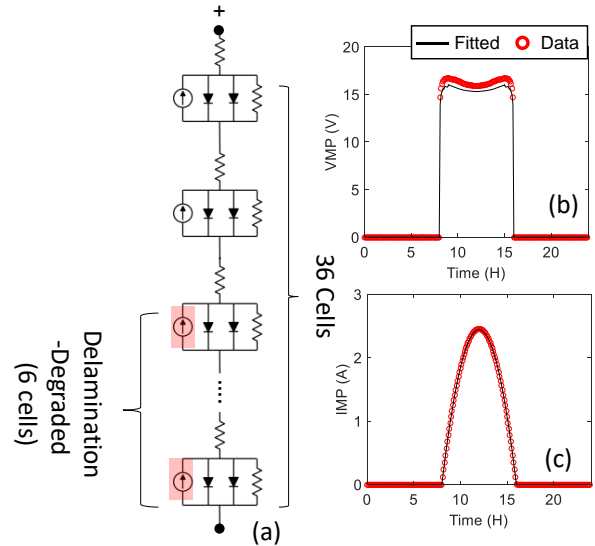

| (d) Extracted parameter by the Suns-Vmp method | | | |
|---|---|---|---|
| | Default (30 cells) | Degraded (6 cells) | Extraction |
| $J_{PH,STC}$ | 282 A/m² | **225 A/m²** | **244 A/m²** |
| $J_{01,STC}$ | 1.3 x 10⁻⁸ A/m² | 1.3 x 10⁻⁸ A/m² | 1.3 x 10⁻⁸ A/m² |
| $J_{02,STC}$ | 4.6 x 10⁻⁴ A/m² | 4.6 x 10⁻⁴ A/m² | 4.6 x 10⁻⁴ A/m² |
| $R_{Sh,STC}$ | 0.12 Ω.m² | 0.12 Ω.m² | 0.12 Ω.m² |
| $R_S$ | 1.7 x 10⁻⁴ Ω.m² | 1.7 x 10⁻⁴ Ω.m² | 1.7 x 10⁻⁴ Ω.m² |

Fig. S5  (a) A schematic of the simulated 36-cell solar module including 6 cells degraded due to delamination. The degraded circuit elements are also highlighted. (b,c) Vmp and Imp of the solar panel using the synthetic weather data in Fig. A1. Circles are simulated data and solid lines are fitting data using the Suns-Vmp method. (d) Table summarizes input parameters (both default and degraded) and extracted parameter set using the Suns-Vmp method (affected parameters are in bold).

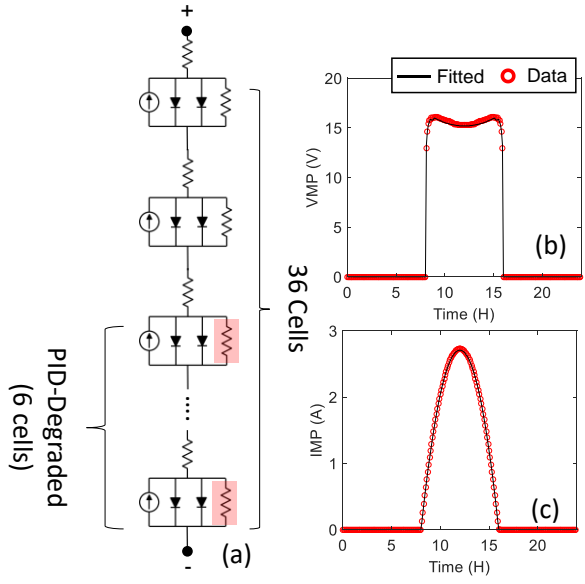

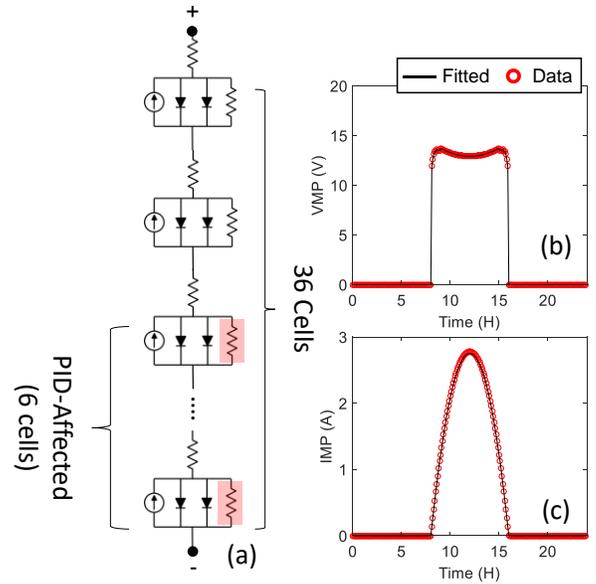

(d) Extracted parameter by the Suns-Vmp method

| | Default (30 cells) | PID- Degraded (6 cells) | Extraction |
|---|---|---|---|
| $J_{PH,STC}$ | 282 A/m² | 282 A/m² | 282 A/m² |
| $J_{01,STC}$ | 1.3 x 10⁻⁸ A/m² | 1.3 x 10⁻⁸ A/m² | 1.3 x 10⁻⁸ A/m² |
| $J_{02,STC}$ | 4.6 x 10⁻⁴ A/m² | 4.6 x 10⁻⁴ A/m² | 4.6 x 10⁻⁴ A/m² |
| $R_{Sh,STC}$ | 0.12 Ω.m² | **0.012 Ω.m²** | **0.026 Ω.m²** |
| $R_S$ | 1.7 x 10⁻⁴ Ω.m² | 1.7 x 10⁻⁴ Ω.m² | 1.7 x 10⁻⁴ Ω.m² |

(d) Extracted parameter by the Suns-Vmp method

| | Default (30 cells) | PID- Degraded (6 cells) | Extraction |
|---|---|---|---|
| $J_{PH,STC}$ | 282 A/m² | 282 A/m² | 282 A/m² |
| $J_{01,STC}$ | 1.3 x 10⁻⁸ A/m² | 1.3 x 10⁻⁸ A/m² | **1.5 x 10⁻⁷ A/m²** |
| $J_{02,STC}$ | 4.6 x 10⁻⁴ A/m² | 4.6 x 10⁻⁴ A/m² | **2.1 x 10⁻² A/m²** |
| $R_{Sh,STC}$ | 0.12 Ω.m² | **0.0012 Ω.m²** | 0.12 Ω.m² |
| $R_S$ | 1.7 x 10⁻⁴ Ω.m² | 1.7 x 10⁻⁴ Ω.m² | **1.8 x 10⁻⁴ Ω.m²** |

Fig. S6 (a) A schematic of the simulated 36-cell solar module including 6 cells degraded due to *moderate* potential induced degradation. The degraded circuit elements are also highlighted. (b,c) Vmp and Imp of the solar panel using the synthetic weather data in Fig. A1. Circles are simulated data and solid lines are fitting data using the Suns-Vmp method. (d) Table summarizes input parameters (both default and degraded) and extracted parameter set using the Suns-Vmp method (affected parameters are in bold).

Fig. S7 (a) A schematic of the simulated 36-cell solar module including 6 cells degraded due to *severe* potential induced degradation. The degraded circuit elements are also highlighted. (b,c) Vmp and Imp of the solar panel using the synthetic weather data in Fig. A1. Circles are simulated data and solid lines are fitting data using the Suns-Vmp method. (d) Table summarizes input parameters (both default and degraded) and extracted parameter set using the Suns-Vmp method (affected parameters are in bold).

## 4. Equations of the Five Parameter Model for Si Solar Modules

Here, we will present the analytical formulation of the five-parameter model [28] used in this paper (see Fig. 2) and the temperature- and illumination- dependency of each parameter in Table S1. Also, detailed description and initial STC value for Siemens M55 [25] of each parameter is listed in Table A2. Note that GSTC = 1000 W/m² and TSTC = 25 °C for standard test condition in for standard test condition in Table S2.

TABLE S1. The equations of the five-parameter model

| Analytical equations for I-V characteristics | |
|---|---|
| $J_{D1} = J_{01}\left(e^{\frac{q(V-JR_S)}{kT}} - 1\right)$ | (A.1) |
| $J_{D2} = J_{01}(e^{\frac{q(V-JR_S)}{2kT}} - 1)$ | (A.2) |
| $J_{Shunt} = \dfrac{(V - JR_S)}{R_{Shunt}}$ | (A.3) |
| $J = J_{PH} + J_{D1} + J_{D2} + J_{Shunt}$ | (A.4) |

| Illumination and temperature dependencies of the parameters | | |
|---|---|---|
| $J_{PH}$ | $J_{PH} = \dfrac{G}{G_{STC}} \times J_{PH,STC} \times (1 + \beta \times (T - T_{STC}))$ | (A.5) |
| $J_{01}$ | $J_{01} = J_{01,STC} \times (\dfrac{T}{T_{STC}})^3 \times \exp(\dfrac{1}{k}(\dfrac{E_{G,STC}}{T_{STC}} - \dfrac{E_G}{T}))$ | (A.6) |
| $J_{02}$ | $J_{02} = J_{02,STC} \times (\dfrac{T}{T_{STC}})^{2.5} \times \exp(\dfrac{2}{k}(\dfrac{E_{G,STC}}{T_{STC}} - \dfrac{E_G}{T}))$ | (A.7) |
| $R_{Shunt}$ | $R_{Shunt} = R_{Shunt,STC} \times \dfrac{G}{G_{STC}}$ | (A.8) |
| $E_G$ | $E_G = E_{G,STC} + \alpha(T - T_{STC})$ | (A.9) |

TABLE S2. Parameter description and their initial STC values for Siemens M55 [6]

| | | |
|---|---|---|
| $J_{SC,STC}$ | Short-circuit current | 282 A/m$^2$ |
| $J_{01,STC}$ | Diode recombination current with ideality factor of 1 | 1.3 x 10$^{-8}$ A/m$^2$ |
| $J_{02,STC}$ | Diode recombination current with ideality factor of 2 | 4.6 x 10$^{-4}$ A/m$^2$ |
| $R_{Sh,STC}$ | Shunt resistance | 0.12 Ω.m$^2$ |
| $R_S$ | Series resistance | 1.7 x 10$^{-4}$ Ω.m$^2$ |
| $\beta$ | temperature coefficient of short-circuit current | 0.49 %/K |
| $E_G$ | Bandgap of Si absorber | 1.12 eV |
| $\alpha$ | temperature coefficient of Si bandgap | -6 x 10$^{-4}$ eV/K |